\documentstyle[aps,multicol,epsf]{revtex}
\begin{document}
\title{  Driven interfaces in disordered media: determination of universality classes from experimental data }
\author{R\'eka Albert, Albert-L\'aszl\'o Barab\'asi$^*$}
\address{Department of
Physics, University of Notre-Dame, Notre-Dame, IN 46556}
\author{ Nathan Carle, Andrew Dougherty}
\address{Department of Physics, Lafayette College, Easton, PA 18042   }
\date{\today}
\maketitle

\begin{abstract}
While there have been important theoretical advances in understanding the universality classes of interfaces moving in porous media, the developed tools cannot be directly applied to experiments. Here we introduce a method that can distinguish the isotropic and directed percolation universality classes from snapshots of the interface profile. We test the method on discrete models whose universality class is well known, and use it to identify the universality class of interfaces obtained in experiments on fluid flow in porous media. 

\end{abstract}
\begin{multicols}{2}
\narrowtext

Interface motion in a random medium presents us with an archetype problem, with direct impact on various phenomena in condensed matter physics, including fluid flow in porous media \cite{fluid flow}, domain growth in disordered magnets \cite{robbins} and flux lines in disordered superconductors \cite{blatter}. In particular, much attention has been focused on understanding the morphological evolution of an interface driven through a disordered medium. Two-phase fluid flow experiments have provided evidence that the morphology of the interface can be either self-similar or self-affine \cite{wong}. The self-similar morphology has been successfully described by various percolation based models, such as invasion percolation\cite{fluid flow}. 
The motion and roughening of self-affine interface morphologies can be quantified by the global interface width
$w(L,t),$
where $L$ is the system size. The study of discrete models and continuum growth equations lead to the observation that the width follows \cite{family,Alb}

\begin{equation}
\label{scaling}
 w(L,t) \sim \left\{\begin{array}{rcl}
t^\beta & \mbox {if}& t \ll t_x\\
L^\alpha & \mbox {if}& t \gg t_x
\end{array}\right.,
\end{equation}
where $\beta$ is the growth exponent and $\alpha$ is the roughness exponent. While both experiments and models have confirmed the applicability of scaling concepts, the obtained roughness exponents are scattered between 0.6 and 1.25,\cite{Alb} questioning the existence of universality, the foundation of the scaling hypothesis (\ref{scaling}).

Motivated by the success of the Kardar-Parisi-Zhang (KPZ) equation in describing interface motion with {\it thermal} noise, it has been proposed that interfaces in porous media are described by the quenched KPZ (QKPZ) equation

\begin{equation}
\label{QKPZ}
\frac {\partial h}{\partial t}=F+ \nu \nabla^2 h+\frac{\lambda}2(\nabla h)^2+\eta({\bf x}, h),    
\end{equation}
where $\eta({\bf x},h)$ represents the quenched noise in the medium. This equation  predicts the existence of a depinning transition: for driving forces $F<F_c$ the interface is pinned, and for $F>F_c$ it moves with a  velocity $v\sim f^\theta$, where $\theta$ is the velocity exponent and $f=(F-F_c)/F_c.$ 

Numerical \cite{alb} and analytical \cite{kardar} studies  indicate that the  discrete models can be grouped into two universality classes depending on the behavior of the QKPZ nonlinearity $\lambda$. Isotropic models (i.e. models that have no growth direction determined by the random medium) have $\lambda=0$ or $\lambda\rightarrow 0$ as $f\rightarrow 0$. The scaling exponents of the {\it isotropic} universality class can be determined analytically from (\ref{QKPZ}) with $\lambda=0$, obtaining $\alpha_i=1$, $\beta_i=0.75$ and $\theta_i=0.33$ in one dimension \cite{nattermann}. However, anisotropy in the medium can induce a $\lambda$ term that diverges at the depinning transition. The exponents characterizing this {\it anisotropic} universality class can be obtained from an exact mapping to directed percolation, providing $\alpha_a=0.63$, $\beta_a=0.63$ and $\theta_a=0.63$ \cite{buldyrev,tang}.

While in numerical simulations the universality class can be identified by measuring $\lambda$ for a tilted interface\cite{alb}, this method cannot be applied experimentally, since a constant tilt cannot be sustained. The scaling exponents obtained for paper wetting \cite{buldyrev,horvath} are in excellent agreement with the prediction of the anisotropic universality class. However, experiments on fluid flow between glass beads provided exponents between $0.6$ and $0.9$ \cite{exp}, not allowing to identify the universality class to which they belong. Thus, in order to properly characterize the experimental results we need to develop methods which, without relying on the determination of the scaling exponents, {\it can identify and distinguish the two universality classes}. Here we make a major step in this direction by introducing a method that can identify the universality class from snapshots of the interface at different time intervals. After demonstrating that the method succesfully identifies the universality class of well known models, we  apply it to identify the universality class of interfaces obtained in experimental investigations. 

Consider an interface $h(i,t)$ moving in a porous medium. Since the interface is rough, we can define a {\it local slope} $s_i$ as being the linear fit to the interface in the region $(i-\delta/2, i+\delta/2)$ (see Fig. 1). If there are slope dependent terms  governing the motion of the interface, (such as the $\lambda(\nabla h)^2$ term in (\ref{QKPZ})), the {\it local velocity} $u_i$ is expected to depend on the local tilt $s_i$.

 The main hypothesis of this paper is that {\it the local slope-dependent velocities satisfy similar scaling laws to the average velocity}. We expect that the local velocity $u(F,s)$, corresponding to the local slope $s$, obeys the scaling law
\begin{equation}
\label {local}
u(F,s)\sim (F-F'_c(s))^{\theta'},
\end{equation}
where $u(F,s)$ is the average local velocity of interface segments with slope $s$, $F_c'(s)$ is the depinning threshold corresponding to these segments, and $\theta'$ is the local velocity exponent.

 In an {\it isotropic medium} there is no selected growth direction, thus the depinning threshold  and the velocity exponent are independent of slope, i.e. $F_c'(s)=F_c$ and $\theta'=\theta$. Consequently, the local velocities have the form
$
u(F,s)=w(s)(F-F_c)^\theta,
$
analogous to the scaling of the average velocity $v(F)=v_0(F-F_c)^\theta,$ i.e. the only slope dependence comes in the non-universal prefactor $w(s)$. Thus we expect that for an isotropic medium the ratio 
\begin{equation}
{\cal V}_i(F,s) \equiv \frac{u(F,s)}{v(F)}=\frac{w(s)}{v_0}
\end{equation}
 is {\it independent of the driving force} $F$.

In contrast, one of the distinguishing features of an {\it anisotropic medium} is that the depinning threshold, $F_c'(s)$, decreases with $s$. Furthermore, it has been shown \cite{kardar} that for globally tilted interfaces (i.e. interfaces which have a nonzero global slope $m$), the velocity exponent is different from $\theta_a=0.63$ obtained for an untilted interface, and it is equal to $1$. Consequently, we cannot exclude the possibility that for anisotropic media $\theta'_a$ defined in (\ref{local}) is different from $\theta_a$. Thus for the anisotropic universality class the ratio ${\cal V}_a \equiv u(F,s)/v(F)$ has a { \it nontrivial $F$-dependence}:
\begin{equation} 
{\cal V}_a(F,s)
=\frac{w(s)(F-F_c'(s))^{\theta'_a}}{v_0(F-F_c)^{\theta_a}}.
\end{equation}

In conclusion, we expect that if ${\cal V}(F,s)$ is independent of the driving force, the interface belongs to the isotropic universality class, while a systematic $F$ dependence is an indication that the interface belongs to the anisotropic universality class. Note that in experiments the driving force is not available, but we can replace it with the  average velocity, because $v(F)\sim(F-F_c)^\theta.$

The advantage of this method is that it can be applied directly to both experimental and numerical data, as long as we have snapshots of the interface at not too distant time intervals. The method is applied as follows: a discretized interface of total length $L$ is partitioned into segments  of length $\delta$. The local slope $s(i)$ for each of the $L/\delta$ segments is determined by fitting a line to it in the region $(i-\delta/2,i+\delta/2)$, where $i=\delta/2,3\delta/2,..,L-\delta/2$. We then repeat the same partitioning for the interface captured at time $t+\tau$, and calculate the local velocities of each of the segments centered at $i$ as $u(i,s)=[h(i,t+\tau)-h(i,t)]/\tau$. As a first step we calculate the average velocity of all segments with slope $s$ when the {\it average velocity} of the interface is $v$,  $u(v,s)=(1/N(s))\sum u(i,s)$, where the sum goes over all segments which have the same slope $s$, and $N(s)$ is the number of such segments (naturally, $\sum_s N(s)=L/\delta$). If $u(v,s)$ plotted as a function of $s$ is a parabola, it indicates that a slope dependent nonlinear term is present in the growth equation. Finally, plotting ${\cal V}(v,s)\equiv u(v,s)/v$ allows us to distinguish the two universality classes: if the $u(v,s)$ curves collapse into a single one, we expect the model to belong to the isotropic universality class, while if a systematic $v$ dependence persists after rescaling, it belongs to the anisotropic universality class. 

To test the applicability of the method, we calculate ${\cal V}(v,s)$ for models whose universality class is known \cite{Alb}, namely the Random Field Ising Model (RFIM) belonging to the isotropic universality class \cite{l}, and the directed percolation depinning (DPD2) model \cite{Alb,tang}, that is known to belong to the anisotropic universality class. These simulations also offer support for the scaling hypothesis (\ref{local}). 

{\it Random Field Ising model---} The definition and the simulation techniques of the RFIM are described in detail in Ref. \cite{alb,koiller}. We have found that the depinning threshold of the local segments is independent of their slope and it is equal to the depinning threshold of the average velocity, $F_c$. Also, we found that the velocity exponent of the local velocities is constant and it is equal to $\theta_i$. Figure 2a shows the local velocities $u(v,s)$ versus the local slope $s$ for different driving forces.  The  parabolic shape of the curves indicates that the KPZ  nonlinearity  is present, in agreement with the results obtained from the global tilt measurements\cite{alb}. As Fig.$\,$2b indicates, the ${\cal V}(v,s)$  curves collapse into a single function, thus confirming that the RFIM belongs to the isotropic universality class. 
 
{\it DPD model---}  The  DPD2 model, introduced by Tang and Leschhorn, is described in detail in Ref.\cite{alb,tang}. As a first step, we investigated the slope-dependence of $F_c'$ and $\theta'$. The scaling of $u(F,s)$ has the following features:

(a) $F_c'(s)$ decreases with $|s|$ obeying the same scaling
law as $F_c(m)$ corresponding to tilted interfaces\cite{kardar};

(b) The  velocity exponent is independent of slope and it has the value $\theta_a'=1$. Consequently, while the average velocity has an exponent $\theta_a=0.63$, the local segments are all tilted from their local hard direction of depinning, and thus they have $\theta_a'=1$\cite{kardar}.

As shown in figure 3a the parabolic shape of the $u(v,s)$ curves indicates the presence of a $\lambda(\nabla h)^2$ term. However, in contrast with the results obtained for the RFIM, there is a systematic shift in the ${\cal V}_a(v,s)$ curves with increasing $v$, confirming that the model belongs to the anisotropic universality class.

{\it Experiments---} The experiments were performed in a thin rectangular plexiglass cell of $125\times 41\times 0.16$ cm, similar to that used in Ref. \cite{exp}, and filled
with unconsolidated glass beads approximately $170 \pm 20 \,\mu$m in
diameter.  The apparatus was continually agitated during filling to help
ensure a reasonably uniform packing of the beads.  Water was injected by a
syringe pump through a series of small holes in a tube aligned along one of
the narrow ends of the cell, approximately mimicking the ideal case of
uniform injection along a line. Images were recorded at $2$-second intervals with a CCD camera and digitized
with a resolution of $512\times 480$ pixels.  Only the central $17$
cm of the system was recorded to avoid edge effects.  The overall scale was
$270\,\mu$m/pixel, or about $1.5$ beads/pixel.  All velocities were measured
in units of bead diameters/second.

Fig.\,4a shows  the local velocity curves obtained from the experimental data. Their shape is consistent with a parabola, indicating the presence of the KPZ nonlinearity\cite {asymmetry}. Fig.\,4b presents ${\cal V}(F,s)$ versus the local slopes for six different  experiments.  While there is some scattering in the data collapse due to the asymmetry of the curves\cite{asymmetry}, in general we can see that there is no systematic shift with the increasing velocity, indicating that while there is a KPZ nonlinearity (suggested by the parabolic shape of ${\cal V}$), the experiments belong to the isotropic universality class.

A convincing summary of our results is shown in Fig.5, where we plot ${\cal V}(v,0)$ as a function of $v$ for the models corresponding to the two universality classes. We see that for the RFIM  ${\cal V}_i(v,0)$ is independent of the average velocity, while for the DPD model, ${\cal V}_a(v,0)$ increases systematically with $v$. Indeed, the driving force dependence of the $s=0$ portions is $(F-F_c)^{\theta'-\theta}$. In the case of the DPD2 model $\theta_a=0.63$ and $\theta_a'=1$, thus ${\cal V}_a(v,0)$ increases with $F-F_c$. On the other hand, ${\cal V}_a(v,0)\sim v^{(1-\theta_a)/\theta_a}$ increases with $v$. On the same plot we show ${\cal V}(v,0)$ determined from the experimental data, which, within the error bars, is consistent with the line corresponding to the RFIM and deviates considerably from the DPD results, underlying our conclusion that the experiments belong to the isotropic universality class. 

Research supported by the NSF Career Award DMR-9710998  and by the sponsors of the Petroleum Research Fund.

\begin{figure}
\caption{ In the interval of length $\delta$ around the position $i$ the interface is fitted with a straight line of slope $s_i$. Assuming that in a consecutive interface $h(i, t+\tau)$ the slope is still $s_i$, the local velocity of point $i$ is $\Delta h/\tau$. 
 }
\label{Fig.1}
\end{figure}

\begin{figure}
\caption{RFIM model. (a) Local velocity $u(v,s)$ versus the local slope $s$ for driving forces $f=(H-H_c)/H_c$ ranging from $0.015$ (bottom curve) to $0.076$ (top curve). The system size is $L=500$ and each result was averaged over $200$ realizations of the disorder. (b) The same curves rescaled by the average velocities.}
\label{RFIM}
\end{figure}

\begin{figure}
\caption{DPD2 model. (a) $u(v,s)$ versus $s$ for driving forces $f=(F-F_c)/F_c$ ranging from $0.016$ (bottom curve) to $0.07$ (top curve). The system size is $L=20000$ and each result was averaged over $500$ realizations of the disorder. (b) The same curves rescaled by the average velocity.}
\label{DPD}
\end{figure}

\begin{figure}
\caption{(a) Dependence of $u(v,s)$ on $s$ for the interfaces obtained in the experiments. The average velocity is given in bead diameters/second. (b) The same curves rescaled by the average velocities.}
\label{EXP}
\end{figure}

\begin{figure}
\caption{Comparison between the behavior of ${\cal V}(v,0)$ with $v$ for the two universality classes. Both variables were rescaled by $v_{min}$. While ${\cal V}_i(v,0)$ is constant, ${\cal V}_a(v,0)$ increases with $v$.  On the same plot we show the values obtained from the experimental interfaces. The error bars for the experimental data were derived from the fluctuations of the average velocity of the interface.}
\label{FINAL}
\end{figure}

\end{multicols}

\begin{references}
\bibitem{asymmetry}
The asymmetry in $u(v,s)$ on Fig.$\,$4 is due to the fact that the experimentally obtained surfaces often were tilted, having an average tilt $m$  corresponding to the position of the minima.
\end{references}
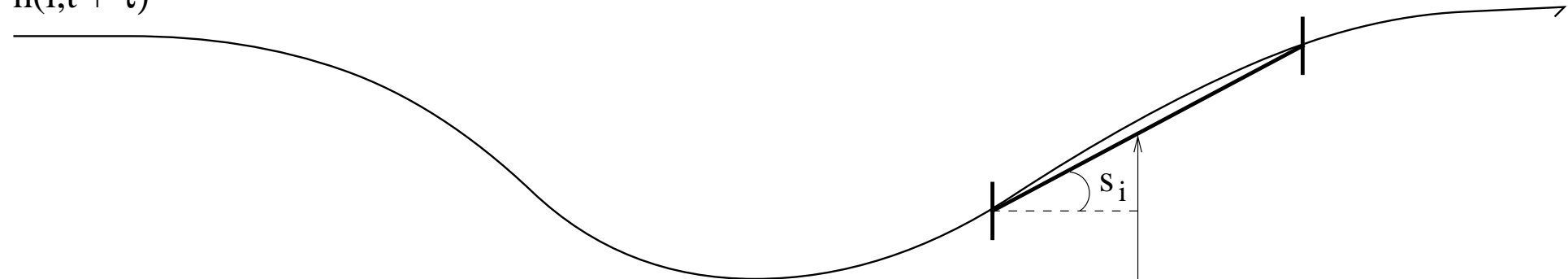
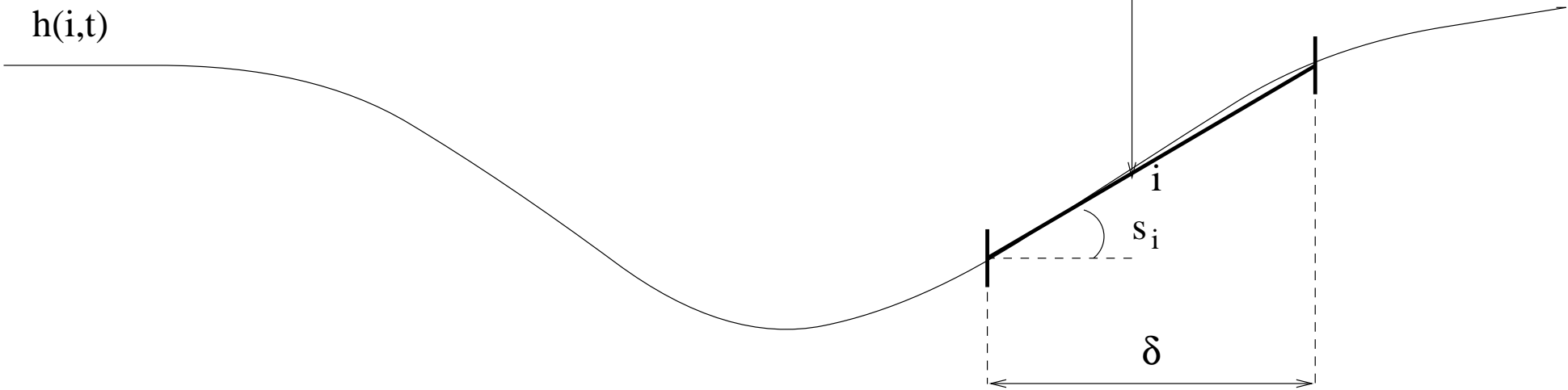
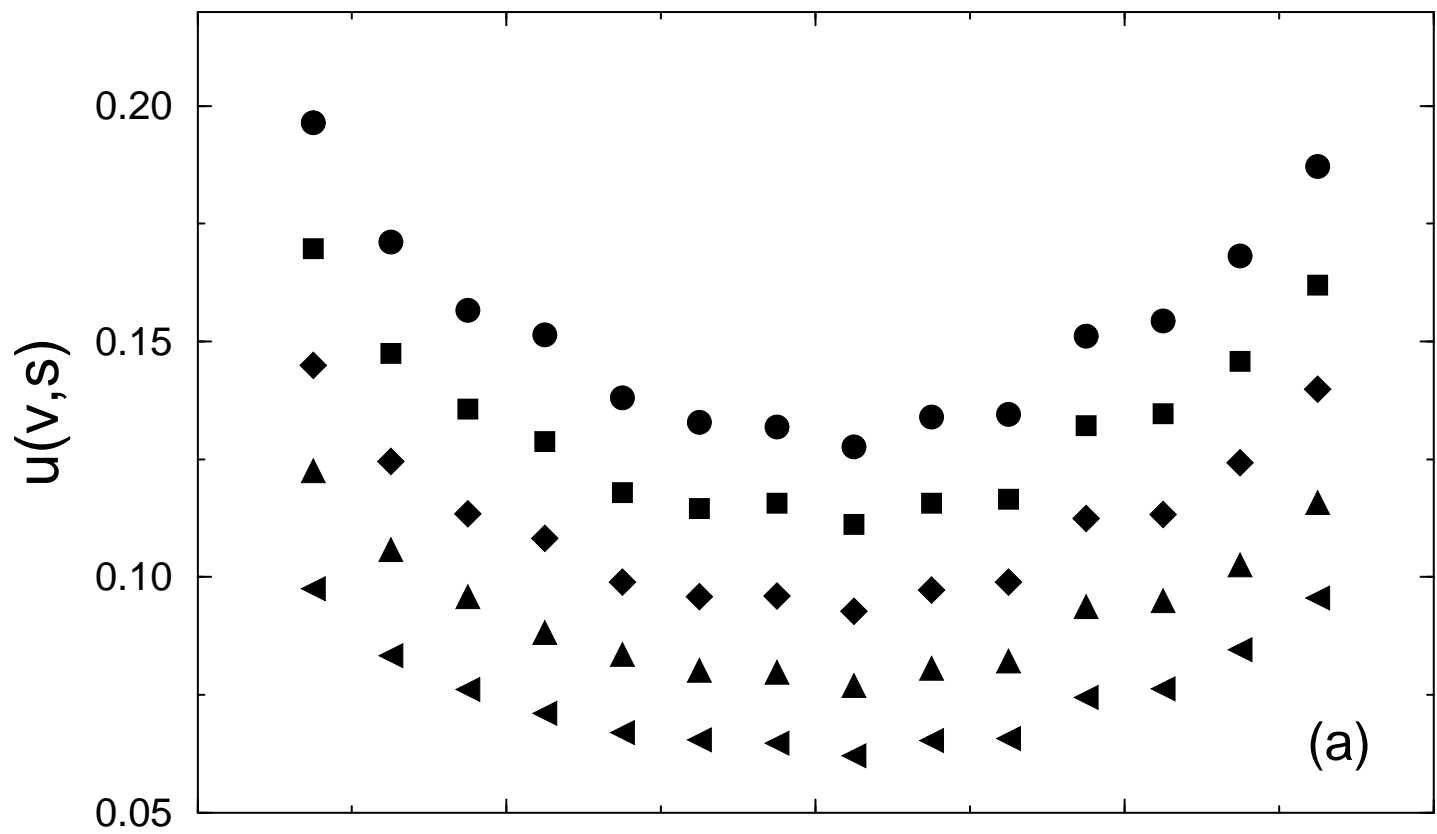
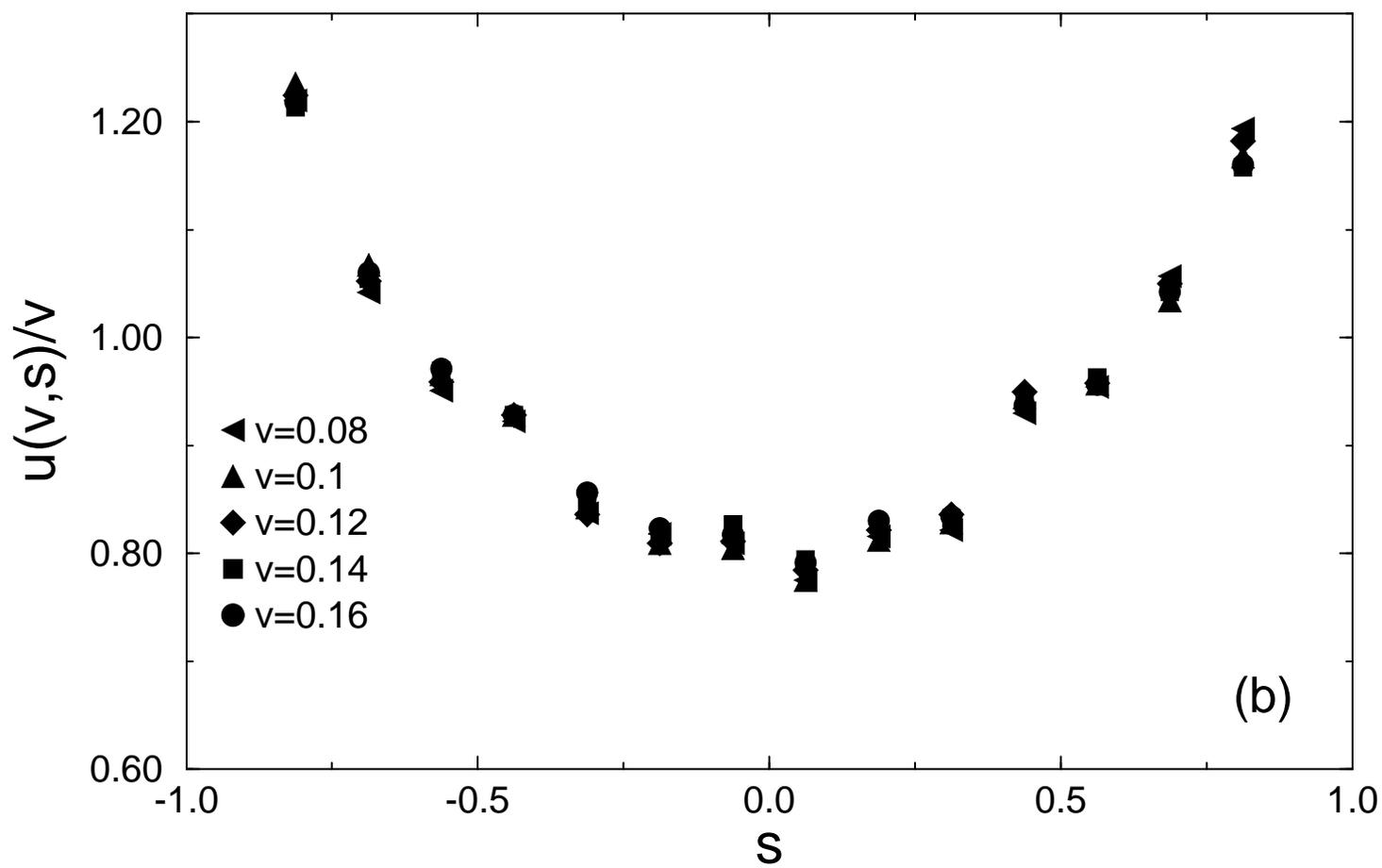
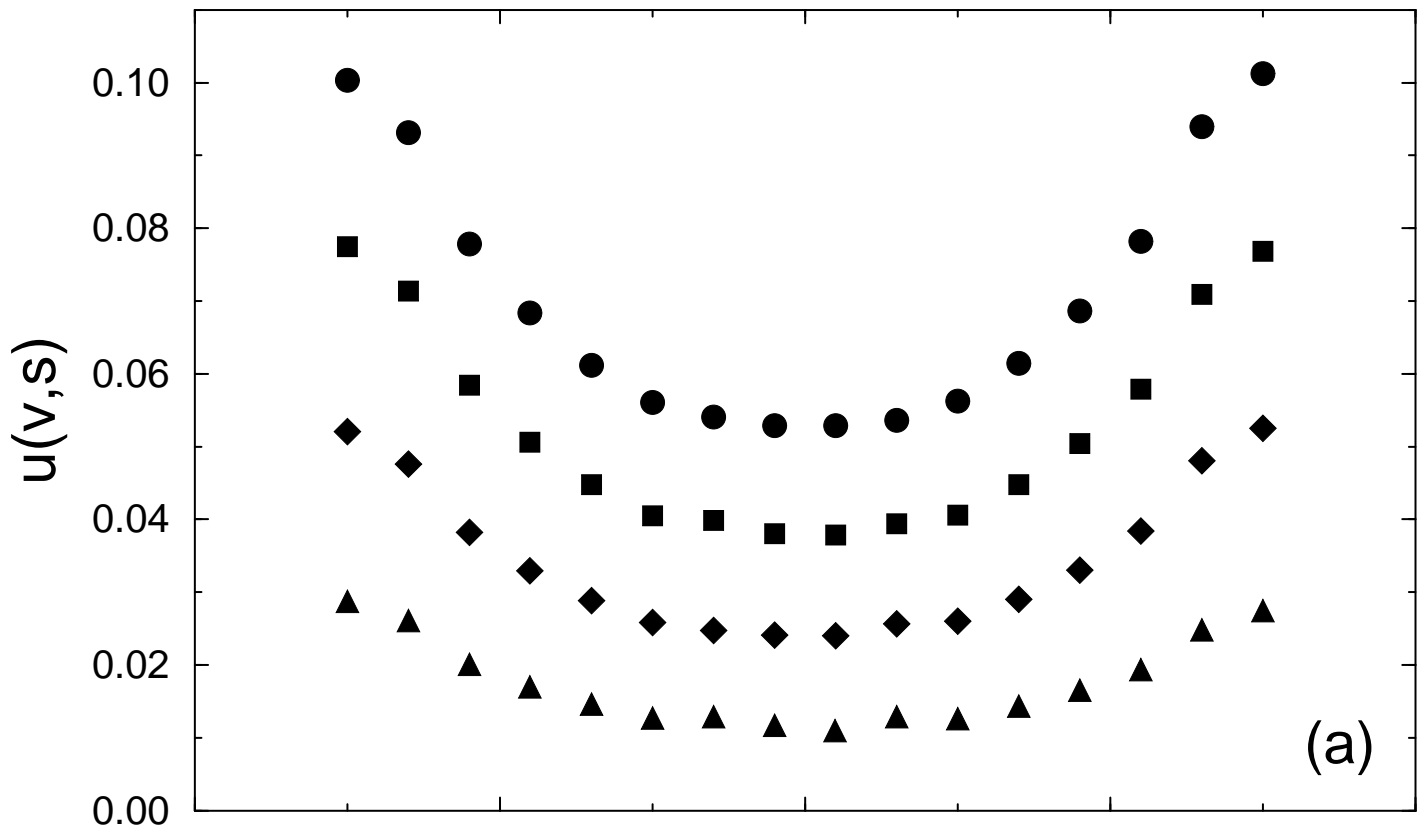
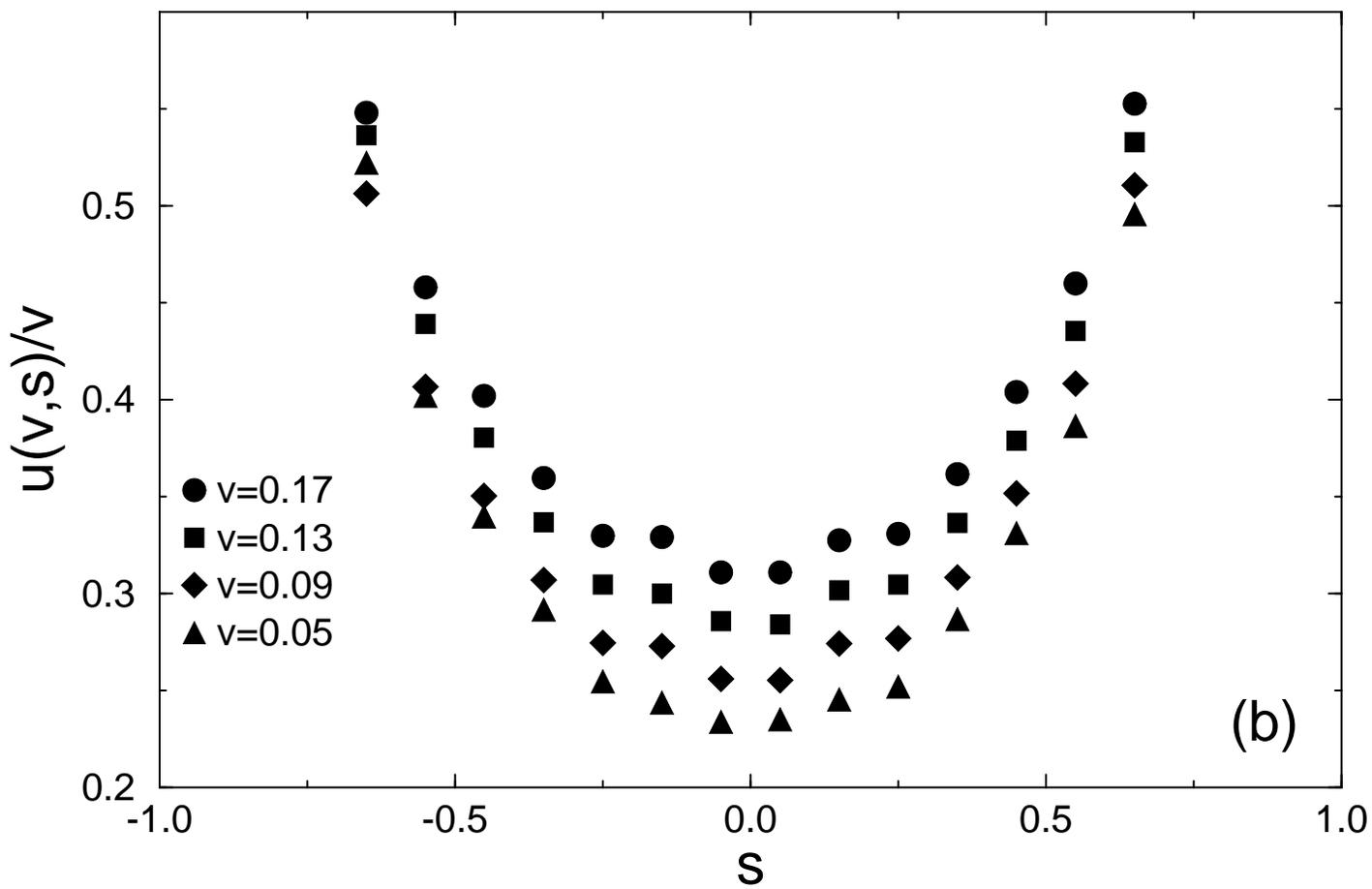
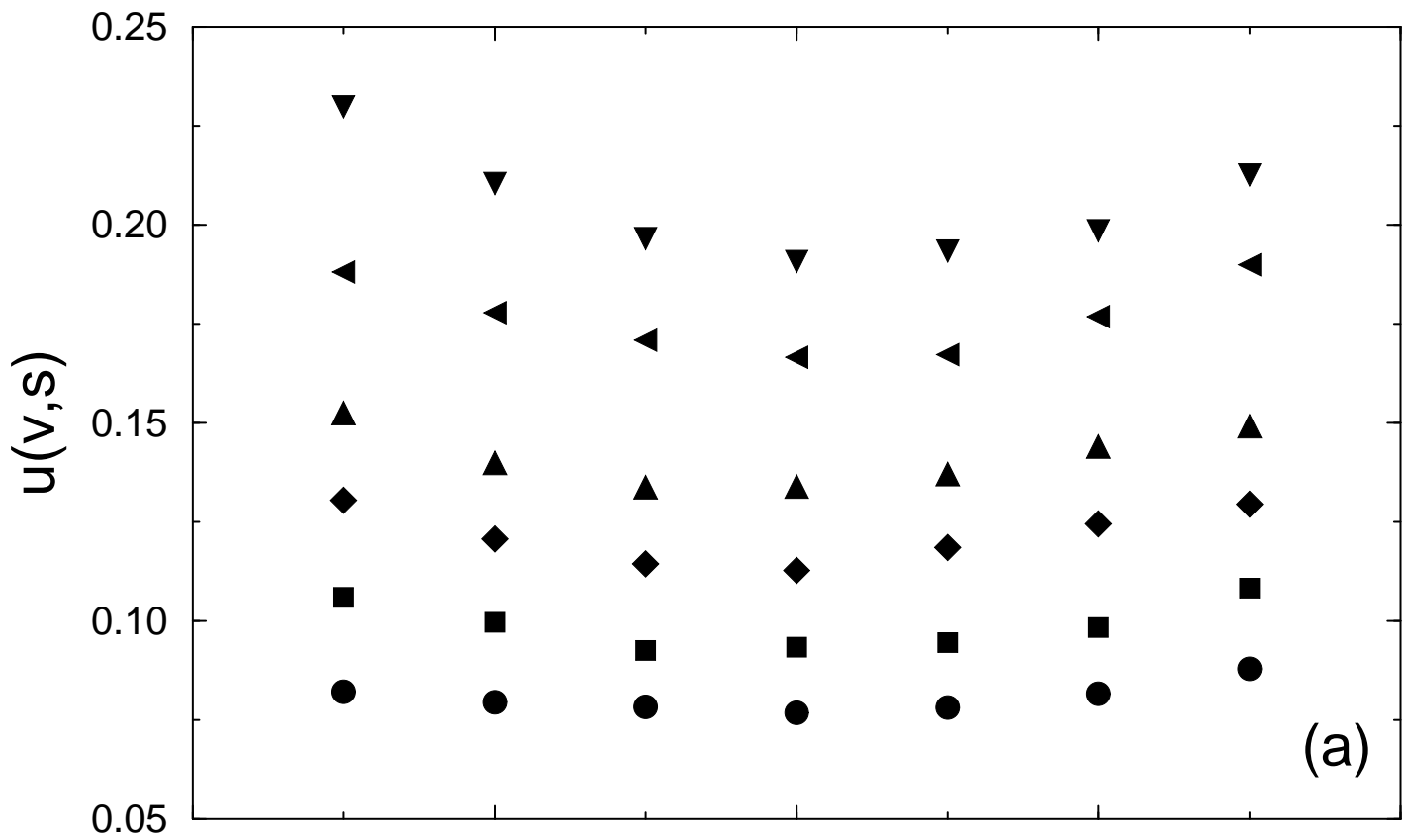
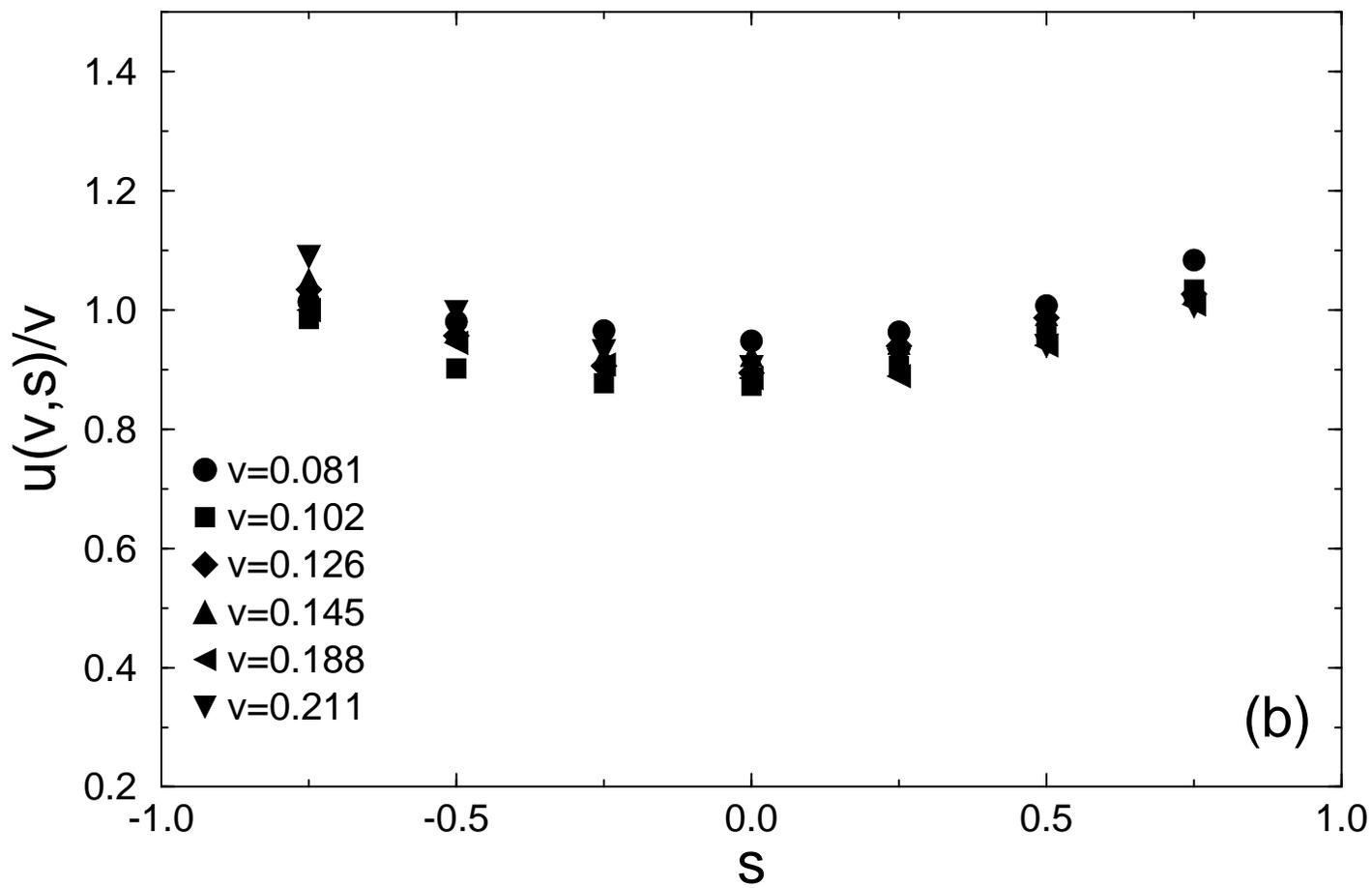
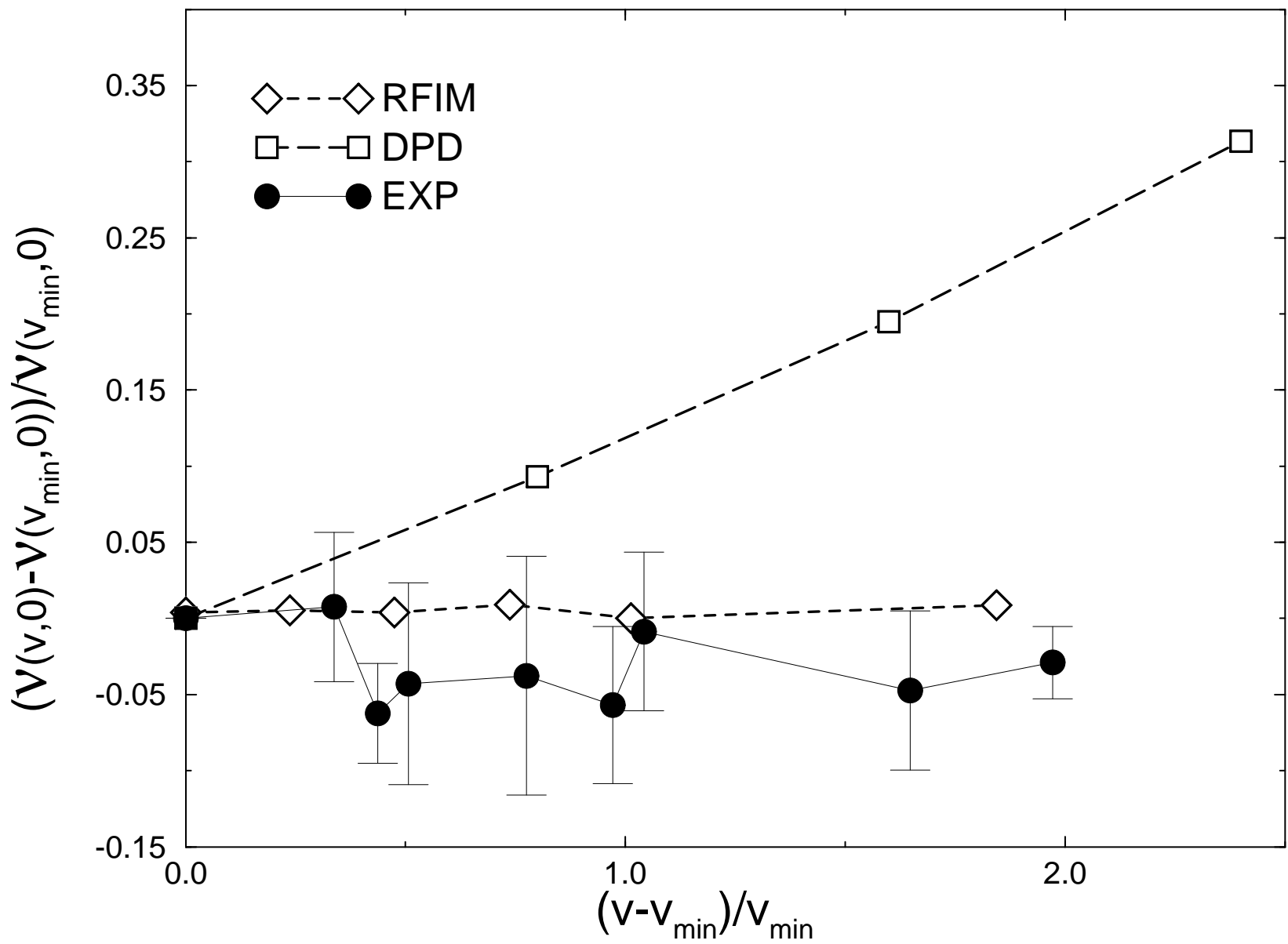
\end{document}